\documentclass[aps,prd,twocolumn,
superscriptaddress]{revtex4}
\usepackage{amssymb,amsmath,graphicx,color,bbold,hyperref}
\begin{document}

\title{On extended symmetries for the Galileon}

\author{Johannes~Noller}
\affiliation{Astrophysics, University of Oxford, DWB,
Keble Road, Oxford, OX1 3RH, UK}

\author{Vishagan~Sivanesan}
\affiliation{UPMC-CNRS, UMR7095, Institut d'Astrophysique de Paris, GReCO, 98bis boulevard Arago, F-75014 Paris, France}

\author{Mikael~von~Strauss}
\affiliation{UPMC-CNRS, UMR7095, Institut d'Astrophysique de Paris, GReCO, 98bis boulevard Arago, F-75014 Paris, France}

\begin{abstract}
We investigate a large class of infinitesimal, but fully nonlinear in the field, transformations of the Galileon and search for extended symmetries. The transformations involve powers of the coordinates $x$ and the field $\pi$ up to any finite order $N$. Up to quadratic order the structure of these symmetry transformations is the unique generalisation of both the infinitesimal version of the standard Galileon shift symmetry as well as a recently discovered infinitesimal extension of this symmetry. 
The only higher-order extensions of this symmetry we recover are (`Galileon dual' versions of) symmetries of the standard kinetic term. 
\end{abstract}

\maketitle

%
%
\newcommand{\be}{\begin{equation}}
\newcommand{\ee}{\end{equation}}
\newcommand{\ba}{\begin{eqnarray}}
\newcommand{\ea}{\end{eqnarray}}

\newcommand{\Lag}{\mathcal{L}}
\newcommand{\td}{\mathrm{d}}
\newcommand{\p}{\partial}
\newcommand{\ph}{\phantom}
\newcommand{\nn}{\nonumber}

\def\bea{\begin{eqnarray}}
\def\eea{\end{eqnarray}}
\def\O{{\cal O}}
\def\L{{\cal L}}
\newcommand{\pa}{\partial}
\newcommand{\comment}[1]{}
\newcommand{\dof}{{\it dof} }
\newcommand{\eoms}{{\it eom}s }
\newcommand{\eom}{{\it eom} }
\def\ba{\begin{eqnarray}}
\def\ea{\end{eqnarray}}
\def\nn{\nonumber}

%
%
\section{Introduction}
Galileon field theories are a class of nonlinear field theories with derivative self-interactions. As their name suggests they satisfy an internal shift symmetry $\pi\rightarrow\pi+C+B_ax^a$ for any constant $C$ and any constant vector $B_a$. They can be defined  as the maximal class of theories with this shift symmetry which still obeys second order field equations. Ever since their recent rediscovery \cite{Nicolis:2008in} they have attracted much attention and these theories have a lot of internal structure yet to be fully explored. Covariant \cite{Deffayet:2009wt} and multi-field generalisations \cite{Deffayet:2010zh} have also been considered but in this work we restrict ourselves to single-field Galileon theories in flat space-time.

The main reason for the interest in these theories is that, despite their nonlinear derivative dependence, they still satisfy second order field equations, thereby avoiding the Ostrogradsky ghost, which generically plagues higher derivative theories (see also~\cite{Fairlie:1991qe}). They are therefore expected to arise in the {\it effective field theory} (EFT) description of many physically interesting situations (see~e.g.~\cite{deRham:2012az} for recent reviews). In particular, they frequently arise in scaling limits of various modified gravity theories exhibiting Vainshtein screening, e.g. in the decoupling limit of brane-world models {\it \`{a} la} DGP \cite{Dvali:2000hr} or in the decoupling limit of nonlinear massive gravity and its generalisations \cite{deRham:2010ik}. For such theories the non-renormalisation theorem of \cite{Luty:2003vm} then ensures that Vainshtein screening can be realised in a controlled fashion within the regime of validity of the EFT, unlike in theories with arbitrary (non-Galileon) irrelevant operators. More generally, Galileons are known to arise in the EFT limit of fluctuating surfaces \cite{deRham:2010eu}. All taken in conjunction, they have a potential interest in many concrete applications all across physics.

In this work we do a systematic search for the existence of extended infinitesimal (but fully nonlinear in the coordinates and field) symmetries of Galileon theories. We find that, up to quadratic order, the symmetry is uniquely fixed to be the standard Galileon shift symmetry plus coordinate/Lorentz transformations together with the quadratic extension recently discovered in~\cite{Hinterbichler:2015pqa}. The only higher order extensions of this symmetry we find are (`Galileon dual' \cite{deRham:2013hsa} versions of) symmetries of the standard kinetic term~\cite{Hinterbichler:2014cwa}.
%
\newline\newline
{\bf Conventions:} We work in $D$ space-time dimensions and frequently employ Einstein summation. We use the notation $\pi_{a}\equiv\p_{a}\pi$, $\pi_{ab}\equiv\p_{a}\p_{b}\pi$ for derivatives of the field $\pi$.

%
\section{Galileon field theories}
The Galileon field theories are defined via an action constructed out of the following Lagrangian,
\be\label{GalL}
\Lag=\sum_{n=1}^D\frac{c_n}{(n+1)}\,\pi\,\pi^{[a_1}_{a_1}\cdots\pi^{a_n]}_{a_n}\,.
\ee
Here and in what follows we have omitted a possible inclusion of a tadpole contribution ($n=0$), which would only change the background solution (i.e.~$c_0=0$ here). The $c_n$ are constant parameters of the theory and the $\pi^{[a_1}_{a_1}\cdots\pi^{a_n]}_{a_n}$ are the completely anti-symmetric products of $\pi^{a}_{b}\equiv\p^{a}\p_{b}\pi$, normalised with unit weight. More explicitly we have
\begin{align}
n&=1:\qquad \pi^{a}_{a}\,,\nn\\
n&=2:\qquad \tfrac1{2}\left(\pi^{a}_{a}\pi^{b}_{b}-\pi^{a}_{b}\pi^{b}_{a}\right)\,,\nn\\
n&=3:\qquad \tfrac1{6}\left(\pi^{a}_{a}\pi^{b}_{b}\pi^{c}_{c}-3\pi^{a}_{b}\pi^{b}_{a}
\pi^{c}_{c}+2\pi^{a}_{b}\pi^{b}_{c}\pi^{c}_{a}\right)\,,\nn\\
\vdots\nn\\
n&=D:\qquad \det(\pi^{a}_{b})\,.
\end{align}
These are, up to normalisation, the unique total derivatives which can be formed out of $\pi^{a}_{b}$ at each order in $\pi$. Due to their anti-symmetric structure, $\pi^{[a_1}_{a_1}\cdots\pi^{a_n]}_{a_n}=0$ for any $n>D$. Note that the $n=1$ term in \eqref{GalL} is the standard kinetic term and, with a mostly plus convention for the metric, the value $c_1=1$ canonically normalises this term in the action.

From \eqref{GalL} the Galileon equations of motion are 
\be\label{GalE}
\mathcal{E}\equiv\sum_{n=1}^Dc_n\,\pi^{[a_1}_{a_1}\cdots\pi^{a_n]}_{a_n}=0\,.
\ee
and we notice that under an infinitesimal transformation, $\pi\rightarrow\pi+\epsilon\delta\pi$, at linear order in $\epsilon$ (i.e.~infinitesimally), the Lagrangian \eqref{GalL} shifts by
\be\label{dL}
\Delta\Lag=\sum_{n=1}^Dc_n\,\delta\pi\,\pi^{[a_1}_{a_1}\cdots\pi^{a_n]}_{a_n}\,.
\ee

%
\section{Symmetries at quadratic order}

Do any Galileon theories exist that are invariant (up to total derivatives) under extensions of the standard Galileon shift symmetry $\pi \to \pi + C + B_a x^a$? The most general transformation for $\pi$ that is a function of $\pi$ itself and coordinates $x^a$ (up to second order in $x^a$ and $\pi$ combined and up to first derivatives acting on $\pi$ -- we will refer to this as {\it quadratic order}) can be written as
\bea \label{2symform}
\delta\pi &=& d_{(0,0)} + d_{(1,0)} b_a^{(1)} x^a +  d_{(0,1)}^{(2)} \pi +   d_{(0,1)} b_a^{(2)} \pi^a \\ \nn &+&  d_{(0,2)}^{(1)} \pi^2 + d_{(0,2)}^{(2)} \pi b_a^{(3)} \pi^a + d_{(1,1)}^{(2)}r_{ab}x^{a}\pi^{b} \\  &+& d_{(2,0)}s_{ab}x^{a}x^{b} + d_{(1,1)}\nn p_{ab}x^{a}\pi^{b}+d_{(0,2)} q_{ab}\pi^{a}\pi^{b}.
\eea
Bracketed indices are labels and all scalar, vector and matrix coefficients $d_{(r,m)}, b_a, s_{ab}, \ldots$ are constant. $s_{ab}, p_{ab}, q_{ab}$ are symmetric, whereas $r_{ab}$ is anti-symmetric.
Plugging the ansatz \eqref{2symform} into the variation \eqref{dL} we compute the contribution to the equations of motion,
\be\label{dE}
\Delta\mathcal{E}=\frac{\p\Delta\Lag}{\p\pi}
-\p_{a}\frac{\p\Delta\Lag}{\p\pi_{a}}+\p_{a}\p_{b}\frac{\p\Delta\Lag}{\p\pi_{ab}}\,.
\ee
Forcing this to vanish provides conditions relating the parameters $c_n$ of the Lagrangian and the parameters $d_{(r,m)}$ of the ansatz \eqref{2symform} and allows us to efficiently find any symmetries.
Doing so (for details see \cite{us}), the most general infinitesimal symmetry transformation up to this order is 
\bea \label{2sym}
\delta\pi &=& d_{(0,0)} + d_{(1,0)} b_a x^a \\  &+& s_{ab} \left( d_{(2,0)} x^{a}x^{b} + \nn d_{(1,1)} x^{a}\pi^{b}+d_{(0,2)} \pi^{a}\pi^{b} \right) \\
\nn &+& d_{(0,1)} b_a^{(2)} \pi^a +  d_{(1,1)}^{(2)} r_{ab}x^{a}\pi^{b},
\eea
where $s_{ab}$ is symmetric {\it and} traceless. All coefficients are free, except $d_{(2,0)}, d_{(1,1)}, d_{(0,2)}$, which have to satisfy
\begin{align}
d_{(0,2)} &= \frac{c_2^2 - c_1 c_3}{c_1^2} d_{(2,0)}, &d_{(1,1)} = \frac{c_2}{c_1} d_{(2,0)}
\end{align}
and, if any of $d_{(2,0)}, d_{(1,1)}, d_{(0,2)}$ are non-zero, we have to restrict to Galileon Lagrangians satisfying
\bea
c_4 = \frac{2 c_2 c_3}{c_1} -\frac{c_2^3}{c_1^2}. 
\eea
\comment{
\begin{align}
d_{(0,2)} &= \frac{c_2^2 - c_1 c_3}{c_1^2} d_{(2,0)} = \frac{c_2^2 - c_1 c_3}{c_1 c_2} d_{(1,1)}, &c_4 = \frac{2 c_2 c_3}{c_1} -\frac{c_2^3}{c_1^2}.
\end{align}
}

The first line in \eqref{2sym} is precisely the standard Galilean shift symmetry, the second line is the non-linear extension recently found by~\cite{Hinterbichler:2015pqa} and the third line is the unique completion of these other symmetries at quadratic order, which simply consists of a coordinate shift and a Lorentz transformation respectively\footnote{We thank Kurt Hinterbichler and Austin Joyce for pointing this out.}. In~\cite{us} we extend this argument to higher orders and also consider terms with higher derivatives acting on $\pi$.

%
\section{Higher order symmetries}

Can this quadratic order symmetry be generalised to higher orders? Extensions involving partially antisymmetric coefficient tensors are somewhat complicated and will be discussed in \cite{us}, but here we conjecture that the general higher order generalisation of the symmetric ($s_{ab}$-dependent) piece of \eqref{2sym} is
\be\label{symform}
\delta\pi=\sum_{(r,m)_{p}}^N\,d_{(r,m)}\,Q_{a_{1}\cdots a_{r}b_{1}\cdots b_{m}}x^{a_1}\cdots x^{a_{r}}\,
\pi^{b_{1}}\cdots\pi^{b_{m}}\,,
\ee
i.e.~a power series in the components of $x^a$ and $\pi^a$, where the $Q$'s are totally symmetric and traceless constant coefficient tensors. The sum in \eqref{symform} runs over all the ordered partitions $(r,m)$ of integers $p=r+m$ (including $0$) up to some $N$ and we define the $Q$'s appropriately whenever $0$ is part of the partition. In general, for each $p$ there are $p+1$ such partitions. This means that the ansatz contains at most $\sum_{p=0}^{N}(p+1)=\tfrac1{2}N(N+3)+1$ arbitrary parameters $d_{(r,m)}$. In order to avoid confusion we stress that we label these partitions according to,
\be
(r,m)=(\#\,\,\mathrm{of}\,\,x^a,\,\#\,\,\mathrm{of}\,\,\pi^a)\,.
\ee 
For example, considering $N=2$ we would sum over the values $p=0,1,2$ with $(r,m)$ taking values in the sets $\{(0,0)\}$, $\{(1,0),(0,1)\}$ and $\{(2,0),(1,1),(0,2)\}$ respectively. Also note that the pattern of generalisation \eqref{symform} is simply to add, at each order in $x^{a}$ and $\pi^{a}$ combined, all possible terms with symmetric and traceless constant coefficient tensors which, at each order, differ at most by an overall constant. We stress that this ansatz is motivated by the above explicit calculation of general symmetry transformations at the lowest orders. 

We now wish to evaluate \eqref{dE} for the conjectured higher order symmetry \eqref{symform}. For this it is convenient to define the {\it traceless symmetric} matrices,
\be\label{Qdef}
^{(r,m)}\mathbb{Q}^{c}_{~d}\equiv x^{a_1}\cdots x^{a_{r}}\,
\pi^{b_{1}}\cdots\pi^{b_{m}}{Q_{a_{1}\cdots a_{r}b_{1}\cdots b_{m}}}^{c}_{~d}\,,
\ee
These matrices satisfy the following useful identities (which hold for any traceless matrix $\mathbb{Q}$),
\be
n\,\mathbb{Q}_{a_1b}\pi^{b[a_1}\pi^{a_2}_{a_2}\cdots\pi^{a_n]}_{a_n}=
-(n+1)\mathbb{Q}^{[a_1}_{a_1}\pi^{a_2}_{a_2}\cdots\pi^{a_{n+1}]}_{a_{n+1}}\,,
\ee
and,
\begin{align}
\mathbb{Q}_{bc}\pi^{bc}\pi^{[a_1}_{a_1}\cdots\pi^{a_n]}_{a_n}=&\,
n\,\mathbb{Q}_{a_1b}\pi^{bc}\pi^{[a_1}_{c}\pi^{a_2}_{a_2}\cdots\pi^{a_n]}_{a_n}\nn\\
&-(n+2)\,\mathbb{Q}^{[a_1}_{a_1}\pi^{a_2}_{a_2}\cdots\pi^{a_{n+2}]}_{a_{n+2}}\,.
\end{align}
These identities can be used to convert all terms appearing in the evaluation of \eqref{dE} into functions of the following form (now there is no non-derivative $\pi$-dependence),
\be
\mathcal{I}_{(n,r,m)}\equiv n\,^{(r,m)}\mathbb{Q}^{[a_{1}}_{a_1}\pi^{a_{2}}_{a_{2}}\cdots\pi^{a_{n}]}_{a_{n}}\,.
\ee
Due to the tracelessness of $\mathbb{Q}$ and the antisymmetric structure we have the very important properties that, independently of the values $(r,m)$,
\be\label{Iprop}
\mathcal{I}_{(1,r,m)}=0\,,\qquad \mathcal{I}_{(n,r,m)}=0\quad\forall n\,>D\,.
\ee
In terms of these functions we find that,
\begin{align}\label{dEsym}
\Delta\mathcal{E} &= \sum_{n=1}^Dc_n\sum_{(r,m)_{p}}^Nd_{(r,m)}\,\Bigl[
m(m-1)\,\mathcal{I}_{(n+2,r,m-2)}\nn \\
&+r(r-1)\,\mathcal{I}_{(n,r-2,m)}
-2rm\,\mathcal{I}_{(n+1,r-1,m-1)} 
\Bigr]\,.
\end{align}
It is now straightforward, for any given $N$, to find the conditions, which need to be satisfied in order to achieve $\Delta\mathcal{E}=0$. Since all the $\mathcal{I}_{(n,r,m)}$ are independent, we simply collect the coefficients of each one of them and demand that they all vanish. This implies the recurrence relation 
\begin{align}\label{c_relGen}
&c_{n}d_{(r,m+2)}(m+1)(m+2)+c_{n+2}d_{(r+2,m)}(r+1)(r+2)\nn\\
&-2c_{n+1}d_{(r+1,m+1)}(r+1)(m+1)=0\,.
\end{align}
This can be solved for any $(r,m)$ but its form is not very illuminating. Before providing an explicit example, some very general remarks can be made by inspection of \eqref{dEsym} and observing the properties \eqref{Iprop}.
\begin{itemize}
\item Any terms with $r=0$ (i.e.~a string of $\pi^{a}$ in \eqref{symform}) leave the $c_{D}$ and $c_{D-1}$ terms invariant. Similarly, any terms with $r=1$ (i.e.~one $x^{a}$ together with a string of $\pi^{a}$ in \eqref{symform}) leave the $c_{D}$ term invariant.
\item Considering the minimum (maximum) value of $2\leq n\leq D$, the coefficients of $\mathcal{I}_{(n_{\mathrm{min}},r-2,m)}$ and $\mathcal{I}_{(n_{\mathrm{max}},r,m-2)}$ in~\eqref{dEsym} have to vanish separately. This implies that the presence of a non-zero $d_{(r,m)}$ with $r\geq2$ ($m\geq2$) requires $c_1\neq0$ ($c_D\neq0$ or $c_{D-1}\neq0$). Since $c_1$ parametrises the standard kinetic term, 
this is also necessary to avoid infinitely strongly coupled solutions.
\end{itemize}

%
\section{A cubic symmetry in $D=4$}
In order to elucidate these points we present the conditions that arise for the $p=2$ and $p=3$ terms in four dimensions, i.e.~considering $N=3$. The ordered partitions fall into the sets $\{(2,0),(1,1),(0,2)\}$ and $\{(3,0),(2,1),(1,2),(0,3)\}$. Computing $\Delta\mathcal{E}$ we find the following set of non-trivial equations,
\begin{align}\label{D4eqs}
\left(c_1d_{(1,1)}-c_2d_{(2,0)}\right)\,\mathcal{I}_{(2,0,0)}&=0\,,\nn\\
\left(c_1d_{(0,2)}-c_2d_{(1,1)}+c_3d_{(2,0)}\right)\,\mathcal{I}_{(3,0,0)}&=0\,,\nn\\
\left(c_2d_{(0,2)}-c_3d_{(1,1)}+c_4d_{(2,0)}\right)\,\mathcal{I}_{(4,0,0)}&=0\,,\nn\\
\left(2c_1d_{(1,2)}-c_2d_{(2,1)}\right)\,\mathcal{I}_{(2,0,1)}&=0\,,\nn\\
\left(2c_1d_{(2,1)}-3c_2d_{(3,0)}\right)\,\mathcal{I}_{(2,1,0)}&=0\,,\nn\\
\left(3c_1d_{(0,3)}-2c_2d_{(1,2)}+c_3d_{(2,1)}\right)\,\mathcal{I}_{(3,0,1)}&=0\,,\nn\\
\left(c_1d_{(1,2)}-2c_2d_{(2,1)}+3c_3d_{(3,0)}\right)\,\mathcal{I}_{(3,1,0)}&=0\,,\nn\\
\left(3c_2d_{(0,3)}-2c_3d_{(1,2)}+c_4d_{(2,1)}\right)\,\mathcal{I}_{(4,0,1)}&=0\,,\nn\\
\left(c_2d_{(1,2)}-2c_3d_{(2,1)}+3c_4d_{(3,0)}\right)\,\mathcal{I}_{(4,1,0)}&=0\,.
\end{align}
It is straightforward to see that the coefficients solving these equations obey the general recurrence relation \eqref{c_relGen}.

A solution to all of the above nine equations (the unique solution for non-zero parameters) is given by,
\begin{align}
c_3&=\frac{3c_2^2}{4c_1}\,, &&c_4=\frac{c_2^3}{2c_1^2}\,,\qquad
d_{(1,1)}=\frac{c_2\,d_{(2,0)}}{c_1}\,,\nn\\
d_{(0,2)}&=\frac{c_2^2\,d_{(2,0)}}{4c_1^2}\,,
&&d_{(2,1)}=\frac{3c_2\,d_{(3,0)}}{2c_1}\,,\nn\\
d_{(1,2)}&=\frac{3c_2^2\,d_{(3,0)}}{4c_1^2}\,,
&&d_{(0,3)}=\frac{c_2^3\,d_{(3,0)}}{8c_1^3}\,.
\end{align}
Note that this leaves two Lagrangian parameters, e.g.~$c_1$ and $c_2$, as well as two parameters, e.g.~$d_{(2,0)}$ and $d_{(3,0)}$, of the transformation undetermined. Furthermore, only the ratio $c_2/c_1$ appears in the $d_{(r,m)}$.

We define $\alpha\equiv c_2/2c_1$ and set $d_{(2,0)}=d_{(3,0)}=1$ by absorbing them into the definition of the corresponding $Q$'s. Ignoring any contributions from (partially) antisymmetric coefficient tensors (like $r_{ab}$), the Galileon theory specified by the above values for the $c_{n}$ has a symmetry up to cubic order in fields and coordinates given by,
\begin{align}\label{dpi3}
\delta\pi=&\,C+B_{a}x^{a}+Q_{a}\pi^{a}
+Q_{ab}x^{a}x^{b}+2\alpha\,Q_{ab}x^{a}\pi^{b}\nn\\
&+\alpha^2\,Q_{ab}\pi^{a}\pi^{b}
+Q_{abc}x^{a}x^{b}x^{c}
+3\alpha\,Q_{abc}x^{a}x^{b}\pi^{c}\nn\\
&+3\alpha^2\,Q_{abc}x^{a}\pi^{b}\pi^{c}+\alpha^3\,Q_{abc}\pi^{a}\pi^{b}\pi^{c}\,.
\end{align}
This result generalises \eqref{2sym} and earlier results by going one order further, but is in some sense trivial. To see this more clearly we discuss the general version of this symmetry.

%
\section{The general form of the symmetry}
An attentive reader may have recognised the binomial coefficients appearing in Eq.~\eqref{dpi3}. This is no accident and, for arbitrary $D$, the higher-order generalisation of \eqref{dpi3} can be written
\be\label{dpiGen1}
\delta\pi=C+B_{a}x^{a}+Q_{a}\pi^{a}+\sum_{m=2}^{N}Q_{a_{1}\cdots a_{m}}\prod_{k=1}^{m}
(x^{a_k}+\alpha\,\pi^{a_k})\,,
\ee
with the $c_n$ obeying $c_{n}=K\,n\,\alpha^{n}$ and where $K,C, B, Q$ are constants as well as $Q_{a_{1}\cdots a_{m}}$ being symmetric and traceless. To prove this, define $Y^{a}\equiv x^{a}+\alpha\,\pi^{a}$ and note that under $\pi\rightarrow\pi+Q_{a_{1}\cdots a_{m}}\prod_{k=1}^{m}Y^{a_{k}}$, for any $m\geq2$, we have
\be
\Delta\mathcal{E}=\sum_{n=1}^{D}c_n\left[
\alpha^2\mathcal{J}_{m}(n+2)-2\alpha\mathcal{J}_{m}(n+1)+\mathcal{J}_{m}(n)\right]\,,
\ee
where we have defined,
\be
\mathcal{J}_{m}(n)\equiv n\,m\,(m-1){Q_{b_{1}\cdots b_{m-2}}}^{[a_{1}}_{a_1}
\pi^{a_{2}}_{a_{2}}\cdots\pi^{a_{n}]}_{a_{n}}\prod_{k=1}^{m-2}Y^{b_k}\,.
\ee
The $\mathcal{J}_{m}(n)$ vanish for $n=1$ and $n>D$. We then find that $\Delta\mathcal{E}=0$ provided that (with $n\geq0$ and $c_0=0$),
\be\label{rec2}
c_{n}\alpha^{2}-2\alpha c_{n+1}+c_{n+2}=0\,\quad\Rightarrow\quad c_{n}\propto n\alpha^n\,.
\ee
This is exactly satisfied by the above values for the $c_{n}$ and the constant of proportionality can be fixed by normalising $c_{1}$. This shows that \eqref{dpiGen1} is indeed an infinitesimal symmetry for arbitrary $D$. 
\comment{
Finally, we note that an alternative way of writing Eq.~\eqref{dpiGen1} is,
\be\label{dpiGen2}
\delta\pi=C+B_{a}x^{a}+Q_{a}\pi^{a}+\sum_{m=2}^{N}Q_{a_{1}\cdots a_{m}}\prod_{k=1}^{m}
\p^{a_{k}}\left(\frac{1}{2}x^2+\alpha\,\pi\right)\,.
\ee
}

What is the nature of this symmetry? Due to the existence of Galileon ``duality" transformations \cite{deRham:2013hsa}, there is a 1-parameter ambiguity. Galileon theories which share this symmetry are therefore unique modulo ``duality" transformations. This can be used to fix the value $c_2=0$, which transforms the Lagrangian with parameters constrained by \eqref{rec2} into a free (non-interacting) Lagrangian with only $c_1\neq0$. Furthermore the symmetry \eqref{dpiGen1} in this case transforms into a string of $x^{a}$, since the duality transformation is essentially a coordinate transformation $x^a \to x^a + \lambda \pi^a$ for a free parameter $\lambda$ \cite{Kampf:2014rka}. This confirms that \eqref{dpiGen1} is indeed a symmetry of the standard kinetic term \cite{Hinterbichler:2014cwa}.

%
\section{Conclusions}
We have performed a systematic search for extended infinitesimal symmetries of the Galileon. At quadratic order we found that the most general such symmetry is given by \eqref{2sym}, which establishes the result found in~\cite{Hinterbichler:2015pqa} as the unique quadratic order extension of the standard Galileon symmetries modulo coordinate and Lorentz transformations. 
Based on this we conjectured a general form for extensions up to any order $N$ in coordinates and field. The parameters $c_n$ of the Lagrangian and the parameters $d_{(r,m)}$ of any such symmetry must obey the recurrence relation \eqref{c_relGen}. Although special cases may be found by studying this recurrence relation, the generic solution for non-zero parameters is uniquely given by \eqref{dpiGen1}. This symmetry is however rather trivial since the parameters of the Lagrangian are constrained such that we are dealing with a  ``dual" version of the free theory and the symmetry reduces to just a string of $x^{a}$, i.e. \eqref{dpiGen1} is the dual version of solely coordinate-dependent symmetries of the kinetic term.
Together with the results of \cite{Cheung:2014dqa}, who found that only one particular Galileon theory has an enhanced soft limit (the quartic Galileon associated with the symmetry found by \cite{Hinterbichler:2015pqa} -- essentially our \eqref{2sym}), this suggests that no `non-trivial' extension of \eqref{2sym} to higher orders exists.
\\

\comment{
We have motivated and presented new infinitesimal symmetries of Galileon theories and highlighted their general features. The unique such symmetries up to quadratic order are given in \eqref{2sym} and a generalisation to arbitrary finite order is given in \eqref{dpiGen1}. Other {\it special} cases may be found by exploring the recurrence relation \eqref{c_relGen}. We emphasise that integrating up the infinitesimal symmetries presented here and in \cite{Hinterbichler:2015pqa} is highly non-trivial and that their `exponentiated' (non-infinitesimal) versions will structurally look very different from the form presented here. 
Also note that \eqref{dpiGen1} in general leaves two parameters of the Lagrangian free. Due to the existence of Galileon ``duality" transformations \cite{deRham:2013hsa}, there is a 1-parameter ambiguity as a result of which Galileon theories which share this symmetry are unique modulo ``duality" transformations. Indeed~\cite{Hinterbichler:2015pqa} found the $N=2$ version of that symmetry (the three first equations of \eqref{D4eqs}) and used the ``duality" (for $D=4$) to set $c_2=0$. It is easy to see, by inspection of \eqref{D4eqs}, that this choice is not consistent for $N>2$. This agrees with the conclusion that the soft scattering behaviour of that particular theory can not be improved beyond cubic order \cite{Cheung:2014dqa,Hinterbichler:2015pqa}. In fact, for $D=4$, the choice $c_{2}=0$ leaves only $c_{1}$ non-zero in the general case and implies that the symmetry is simply a string of $x^{a}$, confirming the observation that this is a symmetry of the standard kinetic term \cite{Hinterbichler:2014cwa}. 

Open questions to explore include associated symmetry algebras and soft scattering behaviour in $D$ dimensions, the relation to enhanced symmetries in modified gravity theories and possible non-linear extensions (both infinitesimal and not) of the symmetry (see~\cite{us} for more).
}

\noindent {\bf Note Added:} The main conclusions of this work changed after private communication with K.~Hinterbichler and A.~Joyce, who pointed out the observation that the higher-order symmetry \eqref{dpiGen1} was the ``dual" version of the string of $x^{a}$ symmetry of the free theory.
\\

%
\acknowledgments
We thank Kurt Hinterbichler and Austin Joyce for helpful comments.
JN acknowledges support from the Royal Commission for the Exhibition of 1851 and BIPAC. The research of VS and MvS leading to these results have received
funding from the European Research Council under
the European Community's Seventh Framework Programme
(FP7/2007-2013 Grant Agreement No. 307934).
%

\end{document}